\documentclass[10pt,printer]{aa}
\usepackage{graphicx}
\usepackage{natbib}
\bibpunct{(}{)}{;}{a}{}{,}
\begin{document}
\title{Observational detection of eclipses of J5 Amalthea by the Galilean satellites}
\author {A.~A.~Christou\inst{1} \and F.~Lewis \inst{2,3,4} \and P.~Roche \inst{2,3,4} \and M.~G.~Hidas \inst{5,6} \and T.~M.~Brown \inst{5,6}
}
\institute{
           Armagh Observatory, College Hill,
           Armagh BT61 9DG, Northern Ireland, UK
           e-mail: aac@arm.ac.uk
      \and Faulkes Telescope Project
           School of Physics and Astronomy, Cardiff University,
           Queens Buildings, 5 The Parade, Cardiff CF24 3AA, UK 
      \and Department of Physics and Astronomy, The Open University, 
           Walton Hall, Milton Keynes MK7 6AA, UK
      \and Division of Earth, Space and Environment, 
           University of Glamorgan, Pontypridd CF37 1DL, UK
      \and Las Cumbres Observatory Global Telescope, 
           6740 Cortona Dr.~Ste.~102, Goleta, CA 93117, USA
      \and Department of Physics, University of California, 
           Santa Barbara, CA 93106, USA
          }
\date{Received 19 April 2010 / Accepted 24 June 2010}
\abstract
{}
{We carried out observations of the small jovian satellite Amalthea (J5) as it was being eclipsed by the Galilean satellites near the 2009 equinox of Jupiter  in order to apply the technique of mutual event photometry to the astrometric determination of this satellite's position.}
{The observations were carried out during the period 06/2009-09/2009 from the island of Maui, Hawaii and Siding Spring, Australia with the 2m Faulkes Telescopes North and South respectively.
We observed in the near-infrared part of the spectrum using a PanStarrs-Z filter with Jupiter near the edge of the field in order to mitigate against the glare from the planet.
Frames were acquired at rates $>1/\mbox{\rm min}$ during eclipse times predicted using recent JPL ephemerides for the satellites. Following subtraction of the sky background from these frames, differential aperture photometry was carried out on Amalthea and a nearby field star.}
{We have obtained three lightcurves which show a clear drop in the flux from Amalthea, indicating that an eclipse took place as predicted. 
These were model-fitted to yield best estimates of the time of maximum flux drop and the impact parameter. These are consistent with Amalthea's ephemeris
but indicate that Amalthea is slightly ahead of, and closer to Jupiter than, its predicted position by approximately half the ephemeris uncertainty
in these directions. We argue that a ground-based campaign of higher-cadence photometry accurate at the $5$\% level or better during the next season of 
eclipses in 2014-15 should yield positions to within $0\farcs05$ and affect a corresponding improvement in Amalthea's ephemeris.}
{}
\keywords{Astrometry -- Eclipses -- Occultations -- Planets and satellites: individual: Jupiter-- Planets and satellites: individual: Amalthea}
\titlerunning{Eclipses of Amalthea by the Galilean satellites}
\authorrunning{Christou, et al.}

\maketitle
\section{Introduction}
The planet Jupiter is attended by four small satellites moving in orbits inside that of Io, the innermost Galilean moon \citep{Thomas.et.al1998}.
Metis, Adrastea and Thebe were discovered during the {\it Voyager} flybys of the 1980s \citep{Jewitt.et.al1979,Synnott1980a,Synnott1980b}. Amalthea, the largest, was discovered 
by E.~E.~Barnard \citep{Barnard1892a,Barnard1892b} during a visual search for new satellites with the 36-in refractor at 
Lick Observatory in California, USA. The innermost pair, Metis and Adrastea, are co-located with Jupiter's main ring system acting as both sources and 
shepherds of ring material. The remaining two, Amalthea and Thebe, fulfil the same function in relation to the so-called Gossamer rings 
\citep{Burns.et.al1999,DePater.et.al1999,OckertBell.et.al1999,DePater.et.al2008,Showalter.et.al2008}. 

The faintness of these inner moons and their proximity to the bright planet
renders them difficult targets for Earth-based observers. Special techniques are often employed to mitigate against the glare from the planet.
These include: utilising methane absorption bands in Jupiter's spectrum \citep{Karkoschka1994}, coronagraphy and post-processing  
\citep{NicholsonMatthews1991,Kulyuk.et.al2002,KulykJockers2004,VeigaVieiraMartins2005,Kulyk2008}. Consequently the uncertainties in 
the ephemerides of these moons are large, ${0}{\farcs}1$-${0}{\farcs}2$ or several hundred km at Jupiter's distance from Earth, 
compared to a few tens of km for the Galilean satellites\footnote{JPL Ephemeris JUP230; http://ssd.jpl.nasa.gov/?sat\_ephem}.
On occasion, higher precision astrometry has been obtained {\it in situ}. For example, satellite positions derived from {\it Cassini} 
ISS NAC images during that spacecraft's flyby of Jupiter in 2000 yielded orbit fits 
with O-C residuals of several tens of km for Adrastea and Metis \citep{EvansPorco2001,Porco.et.al2003} and for Amalthea and Thebe 
\citep{Cooper.et.al2006}. Opportunities to obtain astrometry of this quality require the presence of a spacecraft in the vicinity 
of Jupiter; consequently, they have been quite few in number. 

A powerful ground-based observational technique that can provide very accurate satellite positions over a long period of time and at 
regular intervals is that of mutual event photometry \citep{AksnesFranklin1976,Vasundhara.et.al.2003,Noyelles.et.al.2003,EmelyanovGilbert2006}. There, two 
satellites can be involved in several mutual eclipses or occultations when the planet is at equinox; the observable is a one-dimensional 
time series (a lightcurve) instead of a two dimensional Point Spread Function (PSF). Spatial precision is determined by the rate at which 
the brightness of the satellite can be measured and the relative speed between the two satellites involved in the event, rather than by the seeing.
\citet{AksnesFranklin1978} and \citet{Vachier.et.al2002} have advocated observations of mutual events between Amalthea and the Galilean satellites in order to tie its ephemeris to the significantly more
accurate orbit models of those moons. \citet{Vachier.et.al2002} has pointed out the potential use of an improved ephemeris for Amalthea 
in understanding the dynamics of the jovian ring-moon environment. 

Here we present observations and analyses of photometry of J5 Amalthea obtained as this satellite was being eclipsed by Io (J1) and Europa (J2). 
To our knowledge, these are the first observations of mutual events between Amalthea and the Galilean satellites. 
In the following Section, we describe the strategy adopted in carrying out these observations with particular emphasis on 
mitigating against the glare from Jupiter. In Section 3 we highlight the observing runs that were successfully completed and describe 
the procedure used to process the images from each run. Section 4 describes the model fits to the photometric data while Section 5 relates these fits to 
Amalthea's ephemeris and discusses how our observational strategy may be improved upon. Finally, Section 6 summarises possible avenues for future work.        

\section{Observational Strategy}
Our observations were carried out with the Faulkes Telescopes North (FTN) 
and South (FTS) on the island of Maui, Hawaii and at Siding Spring, Australia respectively \citep[cf Table 1 of][]{Christou.et.al2009}.

In order to plan our observational campaign we generated predictions of such eclipses by incorporating SPICE Ephemeris kernel JUP230 into a prediction program 
previously used in \citet{Christou2005} and \citet{Christou.et.al2009}. We have only considered eclipses of Amalthea by a Galilean satellite, 
assuming that, in other cases, the photometric signature of the mutual event would be lost in the photon noise from the bright ($\Delta m \sim +9$) Galilean. 
Moreover, to obtain a definite signature of an eclipse we have only sought to observe total or near-total eclipses. In fact,
the Galilean satellites would normally saturate the CCD during the actual observations.

All observations were carried out with a PanStarrs-Z filter. This is a broadband filter (FWHM: 104 nm) centred at 870 nm;
as it contains the 890 nm methane absorption feature in Jupiter's spectrum \citep{Karkoschka1994}, it increases the contrast of the satellite against the scattered light from Jupiter. An additional mitigation measure was to offset the telescope pointing so as to place Jupiter outside the field-of-view of the CCD 
keeping the planetary limb 10-15 arcsecond off the edge of the chip. Finally, although use of the batch mode for telescope control would have provided a cadence of 15 sec, 
we opted for the real time mode - where frames were acquired at the lower cadence of 40-50 sec - to maintain the ability to change the observing parameters 
during those challenging observations. Frames were exposed for 5 sec in all cases discussed below.

\section{Observations and data reduction}
Observations were carried out on 2009 June 23; July 21, August 9, 12 and 30; and September 3, 24, 28. 
Acquisition of frames on 23/06 began shortly after 15:30 UT, the predicted start of the eclipse being at 15:41 UT. A JPEG version of the most 
recently obtained frame is posted on the Faulkes Telescope web site within a few minutes of acquisition. 
This allowed near-real-time visual inspection of the telescope pointing, where it 
became apparent that the actual pointing offset was different than the one planned. The centre of the frame was $\sim2.5$ sec in RA 
further to the east bringing most of Jupiter's disk within the field of view. As the eclipse was predicted to start at 15:41 UT it was decided to keep 
the telescope at its current pointing. Later examination of the FITS frames revealed a faint source at Amalthea's predicted position. It should be noted 
that later observations in July and August, otherwise marred by variable weather conditions, showed this telescope pointing offset of $\sim37\arcsec$ to be consistent. The pointing of FTN showed a similarly 
consistent, yet smaller, offset of $\sim1$ sec in RA. We incorporated this into the planning of our observations later in August and September by (a) 
allowing more time (20-30 min) between the start of the observing session and the predicted eclipse in order to ``walk'' the telescope to the desired 
pointing or (b) applying the offset directly into the initial pointing. As a result, good photometry was obtained of two eclipses by Europa on 24/09
and 28/09 using FTS and FTN respectively. The remainder of the paper concentrates on these three observing runs. Their predicted characteristics 
used for planning are given in Table~\ref{tab:pred} while the actual observations are summarised in Table~\ref{tab:obs}.

\begin{table*}
\begin{minipage}[t]{17cm}
\renewcommand{\footnoterule}{}  
\begin{center}
\caption{Predictions for the phenomena observed in this paper.}
\scalebox{1.0}{
\begin{tabular}{ccccccc}\hline\noalign{\smallskip}
Date  &Event & Obs. &   UT Midtime &  Duration \footnote{The interval during which the penumbra is in contact with Amalthea's disk.} & Fractional & Impact Param. \\
(DDMMYY)  & Type & Site & ( HH:MM:SS)  &   (sec) & flux drop & (km)  \\
\noalign{\smallskip}
\hline 
\noalign{\smallskip}                
230609 & 1E5T &  FTS & 15:40:59  & 170  &  1.00 & 18  \\
240909 & 2E5T &  FTS & 11:28:52  & 142  &  1.00 & 696 \\
 280909 & 2E5U &  FTN & 07:01:45  & 209 &  0.89 & 1117 \\
\noalign{\smallskip}
\hline
\end{tabular}}
\label{tab:pred}
\end{center}
\end{minipage}
\end{table*}

\begin{table*}
\begin{minipage}[t]{17cm}
\renewcommand{\footnoterule}{}  
\begin{center}
  \caption{Actual circumstances for the mutual event observations described in this work.}
\scalebox{1.0}{
 \begin{tabular}{cccccccccc}\hline\noalign{\smallskip}
Date   &Event & Obs. &  Exp.~Time &                          & Number of     & UT of       & UT of       &  Seeing &  \\  
(DDMMYY)& Type\footnote{We use the event type notation of \citet{Arlot.et.al2006}.}& Site &    (sec)   &           Filter         & Frames & First  Exp. & Last Exp.   &   ($\arcsec$) & Airmass \\\noalign{\smallskip}\hline \noalign{\smallskip}                                                                                                                      
230609 & 1E5T &  FTS &      5     & PanSTARRS Z &     27        & 15:33:06.4  & 15:56:31.8  & 1.7-2.8 & 1.21 \\ 
240909\footnote{Stars were significantly trailed due to strong winds.} & 2E5T &  FTS &      5     & PanSTARRS Z &     55        & 11:02:04.2  & 11:53:04.5  & 0.8-1.2 &  1.04 \\  
280909 & 2E5U &  FTN &      5     & PanSTARRS Z &     66        & 06:32:09.2  & 07:27:56.4  & 1.2-1.7 &  1.26 \\\hline
\end{tabular}}
 \label{tab:obs}
\end{center}
\end{minipage}
\end{table*}

\begin{figure}[htb]
\centering
\includegraphics[angle=-90,width=8cm]{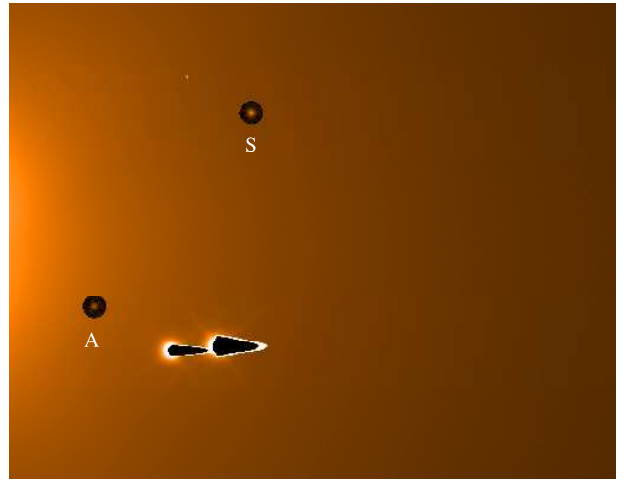}
 \caption[ ]{Subframe of image taken during the observation of the IIEV event on 28/09/2009 from Haleakala, Maui.
The circular apertures indicate the regions of the frame where the background has been subtracted out in advance of
the photometric reduction. Pixel values within the apertures have been multiplied by a factor of 8 for clarity. 
At the centre of each aperture lie Amalthea (``A'') and the star (``S'') used for differential aperture photometry 
of the satellite. Jupiter is off the left edge of the frame which coincides with the edge of the detector. 
The bright sources saturating the detector are, from left to right, Callisto and Europa.  The entire subframe spans $512\times404$ pixels or $143\arcsec\times 113\arcsec$.}
\label{fig:example}
\end{figure}

\subsection{23/06/2009: Io eclipses Amalthea}
The image of Amalthea resided on a significant background of scattered light
from Jupiter. Its gradient is quite severe near the planet and could bias photometric measurements.
To mitigate against this, background pixel values within a circular annulus centred on Amalthea 
and several field stars were used to construct a two-dimensional model of the background 
within the central aperture by performing a third-degree polynomial fit. This was done through the
``Patch'' utility available through the Graphical Astronomy and Image Analysis (GAIA) package available through 
{\it Starlink}\footnote{http://starlink.jach.hawaii.edu/starlink}. The templates resulting from these fits were then 
subtracted from the corresponding apertures leaving ``differenced'' images of the sources resting against a flat, zero-sum, background. 
Fig.~\ref{fig:example} illustrates the result of this procedure for a frame taken on 28/09.
Care was taken not to place either the aperture itself, or the annular region around it used for the fit,
over saturated areas of the frame. In three frames where Amalthea was not clearly visible,
we were guided by their nearest neighbours in placing the aperture.

Extracting the photometry consisted of the following procedure: differenced frames where Amalthea was clearly visible were
used to fit the satellite's motion around Jupiter, modeled as a skewed ellipse on the sky plane. Jupiter's 
motion was also estimated in the fitting process and a nearby star was picked as the point of origin.    
Aperture photometry was then carried out in each frame using the motion model to center the photometric apertures
and sky-estimating annuli. In this case a radius of 2.5 pixels was used for the aperture and
5.5 and 8 pixels respectively for the inner and outer radii of the annulus. These choices minimized the measurement scatter
about the mean. O-C residuals of the fit were 0.67 pixels ($0\farcs19$) and 0.51 pixels ($0\farcs14$) in the horizontal 
and vertical directions respectively. The $1\sigma$ photometric scatter (10\%) was determined by considering only those measurements 
at least a minute before or after the predicted period of the eclipse. The seeing varied between $1\farcs7$ and $2\farcs8$.

\subsection{24/09/2009: Europa eclipses Amalthea}
Images of Amalthea and field stars were distinctly elongated in frames acquired during this observing run. This was likely related to the pointing of 
the telescope near the zenith and the consequent difficulty with which the stepper motors could make subtle adjustments in the pointing of the 
altazimuthally-mounted telescope.
Fitting Amalthea's sky motion yielded O-C residuals of 0.64 and 
0.57 pixels in the horizontal and vertical directions respectively. For the photometry we utilized elliptical apertures with a ratio of 2.0 between the major and the minor axis. Probably for this reason, the photometric scatter was similar to that for the above case (10\%) despite the smaller amount of glare due to
Jupiter being outside the FOV.

\subsection{28/09/2009: Europa eclipses Amalthea}
This representS our highest-quality dataset. The seeing remained between $1\farcs2$ and $1\farcs7$ and the photometric scatter was 6\%, smaller than in 
the other 2 observing runs. Fitting Amalthea's sky motion on the images yielded O-C residuals of $0.43$ and $0.34$ pixels in the horizontal and vertical directions respectively. 

\section{Model fits}
For this stage in the analysis we have used the method of \citet{Christou.et.al2009} to estimate the time $t_{\rm min}$ and distance $b$ of closest 
approach between the two satellites on the impact plane. We note that, unlike the case in that paper, we are measuring the flux from the eclipsed satellite alone; 
hence our data is independent of the satellite albedos. During model fitting we considered the relative velocity $v$ between the eclipsed and eclipsing satellite 
on the impact plane to be fixed and set to the value calculated through the HORIZONS ephemeris service \citep{Giorgini.et.al1996}. Our model cannot currently 
accommodate a satellite with a distinctly non-circular silhouette like Amalthea; instead we have explored the effect of several spherical approximations as explained below.
For reference, \citet{Thomas.et.al1998} reported best-fit triaxal ellipsoid dimensions of $125 \times 79 \times 64$ km. The disks of Io and Europa are assumed 
circular with radii of $1821.6 \pm 0.5$  and $1560.8 \pm 0.5$ km respectively. Amalthea's $1\sigma$ ephemeris uncertainties within JUP230 as reported in 
the JPL Solar System Dynamics website\footnote{http://ssd.jpl.nasa.gov/?sat\_ephem} are 400 km along the direction of motion, 200 km in the radial direction and 100 km in the out-of-orbit plane
direction. Io's and Europa's are 5 km in all directions.

\subsection{23/06/2009: Io eclipses Amalthea}
We have carried out fits to the data assuming circular disks for Amalthea of radius 64, 89 and  125 km. 
The first and third radii correspond to Amalthea's shortest and longest semi-axis respectively while the second radius is $\sqrt{a \times c}$ 
and represents Amalthea as viewed from the Earth near an ansa of its orbit. The differences between these fits are negligible; we adopt a radius of 89 km for all subsequent data fits 
in this paper. The results of all three fits are shown in Table~\ref{tab:fit} and Fig.~\ref{fig:model_fits}. The estimated midtime for the 23/06 event is 13 sec ahead of prediction at 
the $4\sigma$ level of significance. Similarly, the estimated impact parameter is 1050 km higher than the predicted value and significant at the $7\sigma$ level. 
A contour plot of the reduced ${\chi}_{\rm red}^{2}$ goodness-of-fit statistic shows that, in fact, values for $b$ as low as 700 km with corresponding 
$t_{\rm min}$ 6-7 sec ahead of the predictions are within the $1\sigma$ contour. The principal axis of uncertainty within the regions of parameter 
space defined by these contours appears to be inclined at a negative angle to the horizontal, an indication that the two parameters are anticorrelated. Indeed,
the linear correlation coefficient, although small, is negative. This is probably due either (i) to the low cadence of our photometry and 
in particular the lack of data on the egress phase of the eclipse and/or (ii) the relatively large photometric uncertainties.
Even so, the nominal prediction lies outside the 3-sigma contour. At that level of confidence, we can state that (a) the actual value 
of $b$ was higher than 300 km (b) if the value of b is between 300km and 750km then $t_{\rm min}$ was between 15 sec and 0 sec ahead of the prediction. 
\begin{table*}
\begin{minipage}[htb]{17cm}
\renewcommand{\footnoterule}{}  
\begin{center}
  \caption{Result of the two-parameter fits to the lightcurves.}
\scalebox{1.0}{
 \begin{tabular}{cccccccccc}\hline\noalign{\smallskip}
Date     &Event &  UT Midtime & Impact    & Correlation & Minimum  & Mean of & $\sigma$ of  &  $\sigma$ of out-of-eclipse\\
(DDMMYY) & Type &  ( HH:MM:SS.S)       & Parameter (km)  & coefficient ($\rho$) & of $ \chi^{2}_{\mbox{red}}$ & O-C $ (\arcsec)$ &  O-C $(\arcsec)$ & measurements $ (\arcsec)$ \\\noalign{\smallskip}\hline \noalign{\smallskip}                                                                                                                      
230609 & 1E5T & $\mbox{ 15:40:46.2}^{+4.2}_{-4.7}$ & $\mbox{1076}^{+124}_{-150}$ & $-$0.15  & 1.02 & +0.020 & 0.097 & 0.101 \\
240909 & 2E5T & $\mbox{ 11:28:50.1}^{+3.6}_{-4.3}$ & $\mbox{609}^{+266}_{-493}$ & $-$0.22  & 0.32 & $-$0.043 & 0.095 & 0.100 \\  
280909 & 2E5U & $\mbox{ 07:01:39.3} \pm 4.3$ & $\mbox{1109} \pm 60$ & +0.04  & 0.71 & +0.009 & 0.056 & 0.057 \\\hline
\end{tabular}}
 \label{tab:fit}
\end{center}
\end{minipage}
\end{table*}

\subsection{24/09/2009: Europa eclipses Amalthea}
The parameter estimates from the fit to this lightcurve are fairly close - within $1\sigma$ -
to the predicted values. Note that we have chosen not to use the first seven data points in the fit, as they lie $2\sigma$ below Amalthea's 
average out-of-eclipse flux. The goodness-of-fit statistic is somewhat low ($\sim 0.3$), indicating that the data variances have been slightly 
underestimated. Evidence against this hypothesis is that $\chi^{2}_{\rm red}$ decreases if the data variances are artificially increased.
As in the previous case, the parameters are slightly anticorrelated although this is less obvious from the contour plot of the $\chi_{\rm red}^{2}$
quantity. Similarly, the best estimate of $t_{\rm min}$ is slightly ($\sim 2$ sec) ahead of the prediction but, since
the formal uncertainty is twice this difference, this result is not by itself significant. Finally, we note that each of the 
three phases of the eclipse (ingress, totality, egress) are represented in the data, albeit each with a single data point.
  
\subsection{28/09/2009: Europa eclipses Amalthea}
The lightcurve of this event is of higher quality (smaller measurement scatter) than the other two cases. The fitted value of $b$ is relatively well-determined,
perhaps due to the sensitivity of grazing event lightcurves to this parameter, and in very good agreement with the prediction.
The best-fit estimate for $t_{\rm min}$ is ahead of the prediction at a level of significance of between 1 and 2 $\sigma$. 
The two parameters appear to be uncorrelated in this case.  
\section{Discussion} 
\begin{figure*}[t]
\centering
\includegraphics[angle=90,width=18cm]{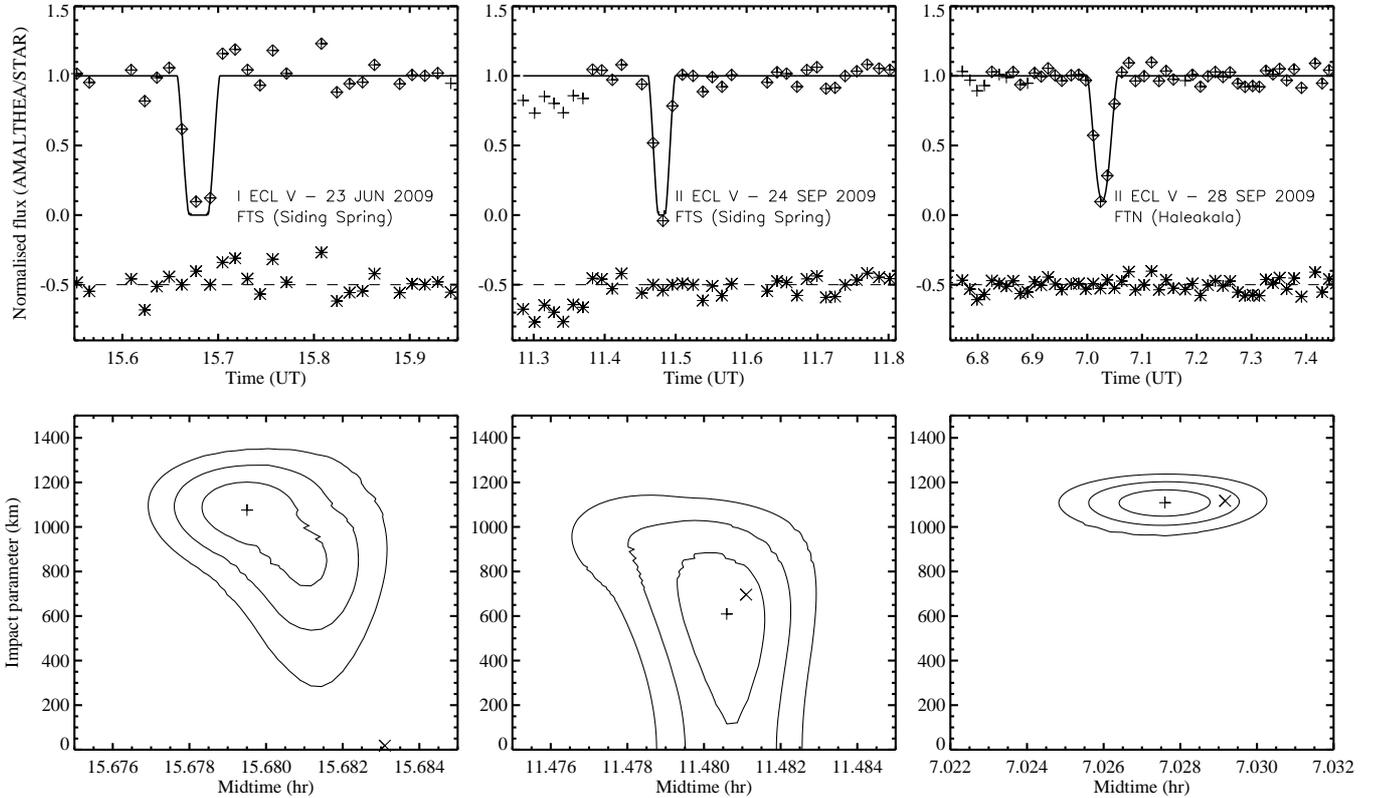}
 \caption[ ]{Model fits to our data. {\it Upper panels}: Best fit models (black curves) to the observations (``+'' signs surrounded by diamonds).
Fit residuals are represented by asterisks. {\it Lower panels}: Parametric contour plots of the reduced chi squared goodness-of-fit statistic ($\chi^{2}_{\rm red}$).
Contours at the $1\sigma$, $2\sigma$, and $3\sigma$ levels are plotted. The best fit value is indicated by a ``+'' sign while the 
ephemeris prediction is indicated by a cross.}
\label{fig:model_fits}
\end{figure*}

The combined results of the three data fits can be used to make statements on the state of Amalthea's ephemeris as inferred from these data
that are stronger than those that can be made from individual lightcurve fits. Firstly, the significantly-higher-than-predicted estimate
of the impact parameter $b$ for the lightcurve obtained on the 23rd June is probably spurious. Arguments in favour of this
conclusion are that (a) the agreement between the $b$ estimates obtained for the other two lightcurves and the predicted values is very good, and 
(b) the large ($\sim$ 1000 km) offset implied for $b$ is not justified given that, for the essentially planar satellite orbits involved here, this
is expected to be comparable to the out-of-plane uncertainty of Amalthea's position ($\sim$ 100 km). This assumption may, in principle, break down if
Amalthea's speed on the impact plane becomes sufficiently small, but that is not the case in any of the three eclipses we observed.
The second major conclusion is that the observed $t_{\rm min}$ is significantly earlier than the prediction. We have constructed a simple model to quantify 
this observation in terms of position offsets in Amalthea's local frame. This frame is defined by the unit vectors $\hat{T}$ (along-track direction),
$\hat{R}$ (radial direction) and $\hat{N}$ (out-of-plane direction). This is, strictly speaking, an orthogonal base vector set only for planar, circular orbits.
We have extracted jovicentric positions and velocities for the Sun, the eclipsing Galilean satellite and Amalthea at the predicted mid-eclipse time,
light-corrected for Jupiter's distance from the Sun, and use those to relate offsets in T, R or N to variations in $t_{\rm min}$ or $b$. This enables us to 
calculate numerically the partial derivatives $\partial  t_{\rm min} / \partial T$, $\partial  t_{\rm min} / \partial R$, $\partial  t_{\rm min} / \partial N$,
$\partial b / \partial T$, $\partial b / \partial R$, $\partial  b / \partial N$. We obtain that the partials of $t_{\rm min}$ ($b$) with respect to N
are orders of magnitude smaller (larger) than the partials in the other two principal directions. This allows us to use $t_{\rm min}$ alone to constrain
likely values of R and T offsets. From the data in Tables \ref{tab:pred} and \ref{tab:fit} for the 24th and 28th September events we obtain
nominal offsets of $T=+175 \pm 219$ km and $R=-77 \pm 95$ km. Further, if we replace one of the equations in this 2x2 linear system with the one
corresponding to the 23rd June event, the two resulting systems yield ranges of $\left[-\mbox{131 km}, \mbox{53 km}\right]$ and $\left[\mbox{44 km}, \mbox{414 km}\right]$ for R and T respectively for a $t_{\rm min}$ offset of $\sim5$ sec. Interestingly, one obtains a similar value if the principal axis of uncertainty in the $\chi_{\rm red}^{2}$ plot 
for that event in Fig~\ref{fig:model_fits} is extended to intersect the $b=0$ axis. We conclude that the uncertainty in our determination of Amalthea's position from 
these observations is 200 km (along-track direction) and 100 km (radial direction) respectively and, furthermore, that Amalthea is ahead of, and slightly closer 
to Jupiter than, its nominal ephemeris location. 
Thus, our observations yield positional uncertainties of one half of the current $1\sigma$ ephemeris uncertainty for Amalthea. The latter also happens to be the 
typical observational uncertainty in modern conventional astrometry of this satellite. The cadence is probably at the limit of what can be employed for useful results, given that the 
confidence in the reported model fits is dependent on the distribution of the photometric points along the lightcurve. \citet{Vachier.et.al2002} recommended a cadence 
of $d/v$ where $d$ is the required spatial resolution of the photometry at Jupiter's distance from the Earth on the plane of sky in km and $v$ is the plane-of-sky velocity 
between the two satellites in km $\mbox{sec}^{-1}$. To sample the mutual events observed here every 200 km, half the uncertainty in conventional astrometry 
($\sim 0\farcs13$), this criterion implies respective cadences of 8, 6 and 11 sec respectively, a factor of 4-8 higher than what was actually used here. 
In general, however, the spatial resolution of the photometric sampling would not map directly to the position uncertainty resulting from the model fit. It would also 
depend eg on the quality of the photometry and the nature of the event that is observed, for example whether it is a total or partial eclipse. Based on this work 
and past experience of the authors with observations of mutual events between the major satellites of Uranus \citep{Christou.et.al2009} we expect that a cadence of 
$\sim 15$ sec and a photometric precision of 5\% would achieve a position determination for Amalthea no worse than $0\farcs050$ (170 km) and perhaps approaching 
the $0\farcs010$ level. For particularly long events where $v$ is of order 10 km $\mbox{sec}^{-1}$ an even lower cadence may be used. Finally, the presence of a 
small but significant correlation between the two parameters in the first two cases, and its absence in the third case, indicates to us that it is linked to the 
uncertainties in the photometric measurement and the low cadence. The above recommended precision of $\leq5$\% and cadence of $\leq15$sec in future observations should mitigate adequately against this.

\section{Implications for future work} 
This work has demonstrated the carrying out of useful mutual event photometry of Amalthea, and possibly other small satellites in the same class,
with instruments that are not specifically fitted out for the purpose of observing faint sources in proximity to the giant planets. That being said,
we found that our low cadence restricted our ability to exploit fully the power of this method to yield high-precision astrometry for this satellite. 
However, this is not an intrinsic limitation of the method or the phenomenon itself; rather a technical feature of the observing apparatus. 
Higher cadence observational setups should allow systematic observations of eclipses of this satellite by the Galileans with apertures smaller 
than the ones use here (eg 1m) starting from the 2014-15 mutual event season. Indeed, sub-meter aperture instruments requiring exposure times of
order a minute could be used on long-duration events. 

 Finally, we point out that an alternative method of accurately measuring the position of a planetary moon through CCD photometry is during 
its ingress in, or egress from, the planet's shadow respectively. Generally, this method requires the precise modelling
of how light is refracted away from the satellite by the planet's atmosphere as well as accounting for albedo variegations across the satellite's surface, phase effects 
and limb darkening \citep{Mallama1991,Mallama1993}. The reported precision of this method as applied to the Galilean satellites \citep{Mallama.et.al2000,Mallama.et.al2010} 
appears to be comparable to the corresponding figures from analysis of mutual event photometry \citep[eg][]{Emelyanov2009}. As eclipses by the planet are 
not limited to the periods around equinox as mutual events are, it is curious that the eclipse method has not met with more widespread use.
One of the culprits may be that, for planetary satellites other than the Galileans, the eclipse condition implies proximity to the planet. 
For Amalthea in particular, where observable eclipses would occur 2-4 times a day, the satellite's distance to the limb of Jupiter at either ingress or egress 
is always $<\mbox{ 10}\arcsec$ whereas it is typically $\sim\mbox{30}\arcsec$ for the observations described in this paper. 
Hence, the scattered light problem will be more severe in the former case, requiring either a more involving reduction procedure to remove it and/or 
the use of specialised equipment such as a coronagraph. Observational data on such eclipses and their astrometric reduction would be useful
in quantifying their suitability for orbit determination work.   

\begin{acknowledgements}
The authors wish to thank the reviewer, Jean-Eudes Arlot, for his comments which improved the manuscript.
The Faulkes Telescope Project is an educational and research arm of LCOGT. FL acknowledges support 
from the Dill Faulkes Educational Trust. Astronomical research at the Armagh Observatory is funded 
by the Northern Ireland Department of Culture, Arts and Leisure (DCAL).
\end{acknowledgements}
\bibliographystyle{aa}
\bibliography{aa_2010_14822}
\end{document}